\documentclass[showpacs,amsmath,amssymb,twocolumn,nofootinbib]{revtex4}
\usepackage{epsfig}
\input{epsf}
\usepackage{amssymb}
\begin{document}
\title{Monodromies and functional determinants in the CFT driven quantum cosmology}
\author{A.O.Barvinsky and D. V. Nesterov}
\affiliation{Theory Department, Lebedev Physics Institute, Leninsky
Prospect 53, 119991 Moscow, Russia}

\begin{abstract}
We apply the monodromy method for the calculation of the functional determinant of a special second order differential operator $\mbox{\boldmath${F}$}=-d^2/d\tau^2+\ddot g/g$, $\ddot g\equiv d^2g/d\tau^2$, subject to periodic boundary conditions with a periodic zero mode $g=g(\tau)$. This operator arises in applications of the early Universe theory and, in particular, determines the one-loop statistical sum for the microcanonical ensemble in cosmology generated by a conformal field theory (CFT). This ensemble realizes the concept of cosmological initial conditions by generalizing the notion of the no-boundary wavefunction of the Universe to the level of a special quasi-thermal state which is dominated by instantons with an oscillating scale factor of their Euclidean Friedmann-Robertson-Walker metric. These oscillations result in the multi-node nature of the zero mode $g(\tau)$ of $\mbox{\boldmath${F}$}$, which is gauged out from its reduced functional determinant by the method of the Faddeev-Popov gauge fixing procedure. The calculation is done for a general case of multiple nodes (roots) within the period range of the Euclidean time $\tau$, thus generalizing the previously known result for the single-node case of one oscillation of the cosmological scale factor. The functional determinant of $\mbox{\boldmath${F}$}$ expresses in terms of the monodromy of its basis function, which is obtained in quadratures as a sum of contributions of time segments connecting neighboring pairs of the zero mode roots within the period range.
\end{abstract}
\maketitle

\section{Introduction}
Successful calculation of quantum corrections on nontrivial backgrounds always presents a challenge and can be accomplished in a closed analytic form only in a limited number of cases. This applies also to calculations in the quantum mechanical sector of field models even despite simplifications occurring in this sector due to its spatial homogeneity or other symmetries. A particular case of such calculations is the class of problems in cosmology involving the differential operator of the form
    \begin{eqnarray}
    \mbox{\boldmath${F}$}=
    -\frac1{g}\frac{d}{d\tau}\,
    g^2\frac{d}{d\tau}\frac1{g}
    =-\frac{d^2}{d\tau^2}+\frac{\ddot g}g,     \label{operator}.
    \end{eqnarray}
where $g=g(\tau)$ is a rather generic function of the cosmic time
depending on the behavior of the cosmological scale factor $a$ and
its time derivative, $\dot a=da/d\tau$. From calculational viewpoint, the virtue of this operator is that $g(\tau)$ represents its explicit basis function -- the solution of the homogeneous equation,
    \begin{eqnarray}
    \mbox{\boldmath${F}$}g(\tau)=0,  \label{equationg}
    \end{eqnarray}
which immediately allows one to construct its second linearly independent solution and explicitly build the Green's function of $\mbox{\boldmath${F}$}$ with appropriate boundary conditions. On the other hand, from physical viewpoint this operator is interesting because it describes long-wavelength perturbations in early Universe cosmology, including the formation of observable CMB spectra \cite{MFB,Mukhanov}, particle creation in cosmology \cite{Mironovetal}, etc.  In particular, for superhorizon cosmological perturbations of small momenta $k^2\ll\ddot g/g$ their evolution operator only slightly differs from (\ref{operator}) by adding $k^2$ to its potential term, whereas in the minisuperspace sector of cosmology, corresponding to spatially constant variables, the operator has exactly the above form.

Up to an overall sign, this operator is the same both in the
Lorentzian and Euclidean signature spacetimes with the time
variables related by  the Wick rotation $\tau=it$. In the Euclidean
case it plays a very important role in the calculation of the
statistical sum for the microcanonical ensemble in cosmology. This ensemble realizes the concept of cosmological initial conditions by generalizing the notion of the no-boundary wavefunction of the Universe \cite{HH} to the level of a special quasi-thermal state and puts it on the basis of a consistent canonical quantization \cite{slih,why}. This concept is very promising from the
viewpoint of the cosmological constant, inflation and dark energy
problems \cite{slih,why,bigboost,tunnel,PIQC}. In particular, it provides a thermal input in the red tilt of the COBE part of the CMB spectrum \cite{DGP/CFT} additional or, perhaps, even alternative to the conventional one generated from primordial vacuum fluctuations in early inflationary Universe \cite{MukhanovChibisov}.

In this statistical theory context, when $\tau$ plays the role of the Euclidean time, the properties of this operator essentially differ
from the Lorentzian dynamics. In the latter case the function $g$ is
a monotonic function of time because of the monotonically growing
cosmological scale factor, whereas in the Euclidean case $g(\tau)$
is periodic just as the scale factor $a(\tau)$ itself and, moreover,
has zeroes at turning points of the Euclidean evolution with $\dot
a=0$, because $g(\tau)\varpropto\dot a(\tau)$. This does not lead to a singular behavior of $\mbox{\boldmath${F}$}$ because $\ddot g$ also vanishes at the zeroes of $g$ \cite{slih,PIQC}, and the potential term of (\ref{operator}) remains analytic (both $g$ and $\ddot g$ simultaneously have a {\em first-order} zero as it follows from the dynamical equation for $a(\tau)$ in \cite{slih}). Nevertheless, the calculation of various quantities associated with this operator becomes cumbersome due to the roots of $g(\tau)$. Among such quantities is the functional determinant of $\mbox{\boldmath${F}$}$ which determines the one-loop contribution to the statistical sum of the CFT driven cosmology of \cite{PIQC}. Since this operator has an obvious zero mode which is the function $g(\tau)$ itself, the functional determinant of $\mbox{\boldmath${F}$}$ should, of course, be understood as calculated on the subspace of its nonzero modes. Thus, the focus of this paper is the calculation of such a {\em restricted} functional determinant of (\ref{operator}), denoted
below by ${\rm Det_*}\mbox{\boldmath${F}$}$. This determinant is
calculated on the space of functions periodic on a compactified
range of the Euclidean time $\tau$ (forming a circle) with its zero
mode $g$ removed or gauged out.

There exist several different methods for restricted functional
determinants. When the whole spectrum of the operator is known this
is just the product of all non-zero eigenvalues. With the knowledge
of only the zero mode, one can use the regularization technique
\cite{McK-Tarlie} or contour integration method \cite{Kirsten-McK,Kirsten-McK1} to extract the regulated
zero-mode eigenvalue from the determinant and subsequently take the
regularization off. Here we use another approach to the definition
of ${\rm Det_*}\mbox{\boldmath${F}$}$ dictated by the gauge-fixing
procedure for the path integral in cosmology \cite{PIQC}. This procedure looks as follows.

As shown in \cite{why}, for a spatially closed cosmology with
$S^3$-topology the microcanonical statistical sum is given
by the path integral over the metric $g_{\mu\nu}$ and matter fields $\phi$ which are periodic on the Euclidean signature spacetime with a compactified time $\tau$ ($S^1\times S^3$ topology). This integral can be transformed by decomposing $(g_{\mu\nu}(x),\phi(x))$ into the minisuperspace  sector of the lapse function $N(\tau)$ and the scale factor $a(\tau)$ of the FRW metric and inhomogeneous ``matter'' fields $\varPhi(x)=(\phi(x),\psi(x),A_\mu(x), h_{\mu\nu}(x),...)$,
    \begin{eqnarray}
    &&Z=
    \!\!\int\!
    D[\,g_{\mu\nu},\phi\,]\;
    e^{-S[\,g_{\mu\nu},\phi\,]}
    =\! \int\!
     D[\,a,N\,]\;
    e^{-\varGamma[\,a,\,N\,]},              \label{1}\\
    &&e^{-\varGamma[\,a,\,N]}
    =\int
    D\varPhi(x)\,
    e^{-S[\,a,\,N;\,\varPhi(x)\,]}\ .             \label{2}
    \end{eqnarray}
Here, $\varGamma[\,a,\,N\,]$ is the Euclidean effective action of
the fields $\varPhi$ (which include also the metric perturbations
$h_{\mu\nu}$) on the FRW background, $ds^2=N^2(\tau)\,d\tau^2+a^2(\tau)\, d^2\Omega^{(3)}$,
$S[\,a,N;\varPhi(x)\,]\equiv S[\,g_{\mu\nu},\phi\,]$ is the
original action rewritten in terms of this minisuperspace
decomposition and $D[\,a,N\,]$ is the canonical measure of integration over periodic histories $\big(N(\tau),a(\tau)\big)$, which incorporates the Faddeev-Popov gauge fixing for the one-dimensional diffeomorphism invariance of $\varGamma[\,a,\,N\,]$.

In the theory with a primordial cosmological constant and a large
number of free (linear) fields conformally coupled to gravity --
conformal field theory (CFT) --  such effective action is dominated
by the contribution of these fields and is exactly calculable with the aid of their local conformal anomaly \cite{FHH,Starobinsky,conf}. The structure of the resulting  action \cite{slih} is
    \begin{eqnarray}
    &&\varGamma[\,a,N\,]=
    \oint d\tau\,N {\cal L}(a,a')+ F(\eta),\,\,
    \eta=\oint d\tau\,\frac{N}a,    \label{action1}
    \end{eqnarray}
where $a'\equiv da/Nd\tau$ and the integration runs over the period
of $\tau$ on the circle $S^1$ of $S^1\times S^3$. Here the effective
Lagrangian of its local part ${\cal L}(a,a')$ includes the classical
Einstein-Hilbert term and the polarization effects of quantum fields and their vacuum (Casimir) energy. A nonlocal part of the action
$F(\eta)$ is the free energy of their quasi-equilibrium excitations
with the effective temperature given by the inverse of the conformal time $\eta$.

The operator (\ref{operator}) arises in the saddle point approximation for the path integral (\ref{1}). It enters the quadratic part of the action (\ref{action1}) in perturbations $\delta a(\tau)$ and $\delta N(\tau)$ on the background of the classical solution $\big(a(\tau),N(\tau)\big)$ for this action. In terms of the new set of perturbation variables
    \begin{eqnarray}
    &&\varphi=\sqrt{|{\cal D}|}\,\delta a,
    \quad n=\frac{\delta N}a
    -\frac{\delta a}{a^2},                \label{variables}\\
    &&{\cal D}\equiv
    \frac{\partial^2\cal
    L}{\partial a'\partial a'}.
    \end{eqnarray}
the part of the action quadratic in $\varphi(\tau)$, which is the canonically normalized perturbation of the cosmological scale factor \cite{PIQC}, reads as
    \begin{eqnarray}
    &&\varGamma_{(2)}[\,\varphi\,]=\frac12\oint
    d\tau\,\varphi(\tau)\mbox{\boldmath${F}$}
    \varphi(\tau),                             \label{action}\\
    &&g=\dot a\,a\sqrt{|{\cal D}|},             \label{g}
    \end{eqnarray}
along with the zero mode $g(\tau)$, which is given as the explicit function of the background history $a(\tau)$ and its derivative $\dot a(\tau)$.\footnote{Here and in what follows we we choose the gauge $N=1$ for the background solution $\big(a(\tau),N(\tau)\big)$, so that time-parametrization invariant derivative $a'\equiv\dot a/N=\dot a$.}

This mode serves as a generator of the global gauge transformation of the gravitational variable $\varphi$, $\delta^\varepsilon\varphi(\tau)\propto g(\tau)\,\varepsilon$.
This transformation is the residual symmetry of the action (\ref{action}) which remains after gauge-fixing the local diffeomorphism invariance of the original action (\ref{action1}) in the gauge $\dot n(\tau)=0$ imposed on the second of the variables (\ref{variables}). Therefore, this symmetry is also subject to the Faddeev-Popov gauge fixing procedure which consists of imposing another gauge $\chi(\varphi)=0$ and inserting in  the path integral the relevant Faddeev-Popov factor. This  gauge condition $\chi(\varphi)$ and the Faddeev-Popov ghost factor $Q/||g||$ can be chosen in the form
    \begin{eqnarray}
    &&\chi=\oint d\tau\,k(\tau)\,\varphi(\tau),
    \quad \delta^\varepsilon\chi
    =\frac{Q}{||g||}\,\varepsilon,          \label{gauge0}\\
    &&Q=\oint d\tau\,k(\tau)\,g(\tau),       \label{Q}\\
    &&||g||^2=\oint d\tau\,g^2(\tau),       \label{norm}
    \end{eqnarray}
where $k(\tau)$ is a gauge fixing function and the role of the generator in the gauge transformation of $\varphi(\tau)$, $\delta^\varepsilon\varphi(\tau)=R(\tau)\,\varepsilon$, is played by the the zero mode $R(\tau)=g(\tau)/||g||$ which is normalized to unity with respect to $L^2$ inner product on a circle.\footnote{In local gauge theories the Faddeev-Popov path integral is not invariant under arbitrary rescalings of gauge generators $R$. Their normalization is always implicitly fixed by the requirement of locality and the unit coefficient of the time-derivative term in the gauge transformation of Lagrange multipliers, $R\varepsilon\sim 1\times\dot\varepsilon+...$, (which follows from the canonical quantization underlying the Hamiltonian version of the Faddeev-Popov path integral). For the residual global symmetry we do not have a counterpart in canonical formalism, and such a founding principle as canonical quantization does not seem to be available. Therefore, we choose this normalization with respect to $L^2$ unit norm corresponding to the canonical normalization of the variable $\varphi$ in (\ref{action}). From the viewpoint of the definition of ${\rm Det}_*\mbox{\boldmath${F}$}$ as the product of operator eigenvalues, this corresponds to the omission of a zero eigenvalue of $\mbox{\boldmath${F}$}$.}

Thus, integration over $\varphi$ takes the form of the Gaussian functional integral with the delta-function type gauge,
    \begin{eqnarray}
    &&\!\!\!\!\!\!\!\!\!({\rm Det}_*{\mbox{\boldmath${F}$}})^{-1/2}\nonumber\\
    && \!\!={\rm const}\!\times\!\!\int D\varphi\,
    \delta\big(\chi\big)\,\frac{Q}{||g||}
    \exp\left(-\frac12\oint
    d\tau\,\varphi\,
    {\mbox{\boldmath${F}$}}\varphi\right),         \label{I}
    \end{eqnarray}
and serves as the definition of the restricted functional determinant of $\mbox{\boldmath${F}$}$. This definition is in fact independent of the choice of gauge by the usual gauge independence mechanism for the Faddeev-Popov integral. In particular, enforcing the gauge $\chi=0$ means that the field $\varphi$ is functionally orthogonal to the gauge fixing function $k(\tau)$ in the $L^2$ metric on a circle, and the above definition is independent of the choice of this gauge fixing function.

The calculation of the restricted functional determinant of (\ref{operator}) as an explicit functional of $g(\tau)$ has already been undertaken in \cite{QCdet} for a particular case of this function $g(\tau)$ having only two zeroes in the period $T$ of the time variable. Here we generalize this single-node result to the multiple nodes case with an arbitrarily high even number $2k$ of roots of $g(\tau)$ within its period range.\footnote{Since a periodic function can have only an even number of roots within its period, the case of the lowest nonvanishing number of roots we will call a {\em single-node} one.} This is the case of the CFT driven cosmology whose statistical sum at different values of the primordial cosmological constant is dominated by the countable set of instantons having $k$ oscillations, $k=1,2,...$, of the scale factor during the Euclidean time period \cite{slih,why} -- the so-called garlands which carry the multi-node zero mode $g(\tau)\propto\dot a$. First we express this determinant in terms of the {\em monodromy} of the second basis function $\psi(\tau)$ of $\mbox{\boldmath${F}$}$, which is linearly independent of $g(\tau)$. In this way we actually reproduce the known monodromy formula of McKane and Tarlie for the restricted functional determinant of the operator subject to periodic boundary conditions \cite{McK-Tarlie,Kirsten-McK,Kirsten-McK1}. Then we derive the answer for this monodromy as an additive sum of contributions of segments of the time variable, $\tau_{i-1}\leq\tau\leq\tau_i$, $i=1,2,...,2k$, separating various pairs of neighboring roots of $g$ on $S^1$, $g(\tau_i)=0$. Each contribution is given by a closed integral expression in terms of $g(\tau)$ on an underlying segment. Finally, we reproduce the previously obtained result for a single-node case of \cite{QCdet} and accomplish the paper with conclusions.

\section{Setting the problem and summary of results}

Let us start with the setting of the problem and a brief formulation of the final result. The operator (\ref{operator}) is defined on a circle range of the time variable $S^1$ having the circumferance $T$, which is parameterized by $\tau$
    \begin{eqnarray}
    \tau_0<\tau<\tau_0+T,        \label{circle}
    \end{eqnarray}
with the points $\tau_0$ and $\tau_0+T$ being identified. This range can be infinitely extended to the whole axis $-\infty<\tau<\infty$, multiple covering of $S^1$, on which the function $g(\tau)$ and, consequently, the operator $\mbox{\boldmath${F}$}$ are periodic with the period $T$,
    \begin{eqnarray}
    g(\tau+T)=g(\tau).        \label{periodicg}
    \end{eqnarray}
In the fundamental domain this periodic function has $2k$ simple roots
    \begin{eqnarray}
    &&\tau_0<\tau_1<\tau_2<...<
    \tau_{2k}=\tau_0+T,          \label{roots}\\
    &&g(\tau_i)=0,\quad
    \dot g(\tau_i)\neq 0,      \label{simpleroots}\\
    &&\ddot g(\tau_i)=0,       \label{ddotg}
    \end{eqnarray}
one of them coinciding with the final (or starting) point of this domain. Another important assumption is that the second order derivative of this function at its roots is vanishing, which will be important for analyticity properties of our formalism.

For any two functions $\varphi_1$ and $\varphi_2$ the operator $\mbox{\boldmath${F}$}$ determines their Wronskian $W[\varphi_1,\varphi_2]$ and Wronskian relation defined by the equations
    \begin{eqnarray}
    &&W[\varphi_1,\varphi_2]\equiv \varphi_1\dot\varphi_2-\dot\varphi_1\varphi_2,\\
    &&\int\limits_{\tau_-}^{\tau_+} d\tau \,\varphi_1 \overrightarrow{\mbox{\boldmath$F$}} \varphi_2
    =\int\limits_{\tau_-}^{\tau_+} d\tau \,\varphi_1 \overleftarrow{\mbox{\boldmath$F$}}
    \varphi_2 - W[\varphi_1,\varphi_2]\,
    \Big|^{\;\tau_+}_{\;\tau_-}.        \label{Wronskianrelation}
    \end{eqnarray}
Arrows here denote the direction of action of the operator $\mbox{\boldmath$F$}$, i. e. $\varphi_1\! \overleftarrow{\mbox{\boldmath$F$}}=(\mbox{\boldmath$F$}\varphi_1)$, and the Wronskians appear as total derivative terms generated by integration by parts of the derivatives in $\mbox{\boldmath$F$}$. When both $\varphi_1$ and $\varphi_2$ satisfy a homogeneous equation with the operator $\mbox{\boldmath${F}$}$, their Wronskian turns out to be constant. Also the vanishing Wronskian implies linear dependence of these solutions.

Let us now consider the solution of the homogeneous equation $\psi(\tau)$ normalized by a unit Wronskian with $g$
    \begin{eqnarray}
    &&\mbox{\boldmath${F}$}\psi(\tau)=0,   \label{equationpsi}\\
    &&W[g,\psi]=1
    \end{eqnarray}
Together with $g(\tau)$ this solution forms a set of linearly independent basis functions of $\mbox{\boldmath$F$}$. However, in contrast to $g(\tau)$ the basis function is not periodic, because we assume that the operator (\ref{operator}) has only one periodic zero mode smoothly defined on a circle (\ref{circle}). On the other hand,
when considered on the full axis of $\tau$, due to periodicity of $g(\tau)$ this operator is also periodic $\mbox{\boldmath$F$}(\tau+T)=\mbox{\boldmath$F$}(\tau)$. Therefore $\psi(\tau+T)$ is also a solution of the equation $\mbox{\boldmath$F$}(\tau)\psi(\tau+T)=0$, and consequently it can be decomposed into a linear combination of the original two basis functions with constant coefficients
    \begin{eqnarray}
    \psi(\tau+T)=\psi(\tau)+\Delta\, g(\tau).     \label{Monpsi}
    \end{eqnarray}
The unit coefficient in the first term follows from the conservation in time of the Wronskian of any two solutions of the equation (\ref{equationpsi}), periodicity of $g(\tau)$ and an obvious fact that $W[g,g]=0$, $1=W[g(\tau+T),\psi(\tau+T)]=W[g(\tau),\psi(\tau) +\Delta\, g(\tau)]$. The coefficient $\Delta$ in the second term is nontrivial -- this is the {\em monodromy} of $\psi(\tau)$ which will play a central role in the construction of the determinant.

Below we show that in terms of this monodromy and the square of the norm of the zero mode (\ref{norm}) the restricted functional determinant of the operator $\mbox{\boldmath$F$}$ subject to periodic boundary conditions has a very simple form
    \begin{eqnarray}
    &&{\rm Det_*}\,
    \mbox{\boldmath${F}$}
    =\Delta\oint d\tau\,g^2(\tau).  \label{det}
    \end{eqnarray}
In fact, this is the McKane-Tarlie formula (Eq.(5.2) of \cite{McK-Tarlie}) obtained by the regularization and contour integration methods \cite{McK-Tarlie,Kirsten-McK,Kirsten-McK1} based on the earlier work of Forman \cite{Forman} for a generic second order differential operator. We reproduce this formula by the variational method for functional determinants. Beyond this, the structure of the operator (\ref{operator}) makes the problem exactly solvable and allows one to find the monodromy (\ref{Monpsi}) in quadratures as an explicit functional of $g(\tau)$. The answer reads in terms of the set of solutions of (\ref{equationpsi}) explicitly defined by the following integrals for $i=1,...,2k$,
    \begin{eqnarray}
     &&\varPsi_{i}(\tau)= g(\tau)
     \int\limits_{\tau_{i}^*}^{\tau}
     \frac{dy}{g^2(y)},\;
     \tau_{i-1}<\tau,\tau_{i}^*<\tau_i,              \label{Psis}
    \end{eqnarray}
on the segments of $\tau$-range connecting the pairs of neighboring roots of $g(\tau)$. Here $\tau_i^*$ are the auxiliary points arbitrarily chosen in the same segments, and all these solutions are normalized by the unit Wronskian with $g(\tau)$, $W[g,\varPsi_i]=1$. Then the monodromy in question reads as the additive sum of contributions of all these segments
    \begin{eqnarray}
    &&\Delta=\sum\limits_{i=1}^{2k}
    \varDelta_i,                     \label{Delta}\\
    &&\varDelta_i=
    \frac{\dot\varPsi_i(\tau_i)}{\dot g(\tau_i)}
    -\frac{\dot\varPsi_i(\tau_{i-1})}
    {\dot g(\tau_{i-1})}\nonumber\\
    &&\qquad=-\Big(\varPsi_i(\tau_i)\,
    \dot\varPsi_i(\tau_i)
    -\varPsi_i(\tau_{i-1})\,
    \dot\varPsi_i(\tau_{i-1})\Big).                 \label{Deltai}
    \end{eqnarray}

The main property of these functions $\varPsi_i(\tau)$ is that each of them is defined in the $i$-th segment of the full period of $\tau$ where the integral (\ref{Psis}) is convergent because the roots of
$g(\tau)$ do not occur in the integration range. Its limits are well defined also at the boundaries of this segment,
    \begin{eqnarray}
    \varPsi_i(\tau_{i-1})=
    -\frac1{\dot{g}(\tau_{i-1})},\quad
    \varPsi_i(\tau_i)
    =-\frac1{\dot{g}(\tau_i)},    \label{psilimits}
    \end{eqnarray}
because the factor $g(\tau)$ tending to zero compensates for the
divergence of the integral at $\tau\to\tau_i-0$ and $\tau\to\tau_{i-1}+0$. Moreover, because of (\ref{ddotg}) the functions $\varPsi_i(\tau)$ are differentiable in these limits, and all the quantities which enter the algorithm (\ref{Deltai}) are well defined. In particular, for any such time segment $[\tau_{i-1},\tau_i]\equiv[\tau_-,\tau_+]$ the derivatives of $\varPsi(\tau)$ at its boundaries are given by the convergent integral (we omit the label $i$ marking any of the functions $\varPsi_i(\tau)$ and their auxiliary points $\tau_i^*$ in (\ref{Deltai}))
    \begin{eqnarray}
    \dot\varPsi(\tau_\pm)=
    \int\limits_{\tau^*}^{\tau_\pm} dy\,
    \frac{\dot g(\tau_\pm)
    -\dot g(y)}{g^2(y)}+\frac1{g(\tau^*)}.    \label{derivative}
    \end{eqnarray}
Note that the integrand here is finite at $y\to\tau_\pm$ because of $\ddot g(\tau_\pm)=0$. These properties of $\varPsi_i(\tau)$ guarantee that the obtained result is independent of the choice of the auxiliary point $\tau_i^*$ for each $\Delta_i$, and the monodromy (\ref{Delta}) is uniquely defined.

Final comment concerns the overall normalization in (\ref{det}). The formula of McKane-Tarlie \cite{McK-Tarlie,Kirsten-McK,Kirsten-McK1} or the variational method which we use below, in principle, give only the ratio of determinants for two different operators (say with different functions $g$ and periods $T$), whereas each determinant contains an infinite numerical factor generated by UV divergent product of eigenvalues of $\mbox{\boldmath${F}$}$. One can only say that this factor is independent of $g(\tau)$ and $T$ and depends on the UV regularization. Here we fix it by the zeta-function regularization in which it turns out to be equal to one.

\section{Variational expression for the determinant}
Representing the delta function of the gauge condition in (\ref{I})
via the integral over the Lagrangian multiplier $\pi$ we get the
Gaussian path integral over the periodic function $\varphi(\tau)$
and the numerical variable $\pi$,
    \begin{eqnarray}
    &&({\rm Det}_*{\mbox{\boldmath${F}$}})^{-1/2}\nonumber\\
    &&\qquad\qquad={\rm const}\times \frac{Q}{||g||}\int D\varphi\,d\pi\,\exp\Big(-
    S_{\rm eff}[\,\varphi(\tau),\pi\,]\,\Big)\nonumber\\
    &&\qquad\qquad={\rm const}\times Q\,||g||^{-1}\Big({\rm Det}\,
    \mathbb{F}\Big)^{-1/2}.                         \label{I2}
    \end{eqnarray}
Here $S_{\rm eff}[\,\varphi(\tau),\pi\,]$ is the effective action of
these variables and $\mathbb{F}$ is the matrix valued Hessian of
this action with respect to $\varPhi=(\varphi(\tau),\pi)$,
    \begin{eqnarray}
    &&S_{\rm eff}[\,\varphi(\tau),\pi\,]=
    \oint
    d\tau\,\Big(\,\frac12\,\varphi
    {\mbox{\boldmath${F}$}}\varphi-i\pi k\varphi\Big),\\
    &&\mathbb{F}=
    \frac{\delta^2S_{\rm eff}}
    {\delta\varPhi\, \delta\varPhi'}=
    \left[\,\begin{array}{cc} \;{\mbox{\boldmath${F}$}}\,
    \delta(\tau,\tau')&\,\,\, -i k(\tau)\,\\
    &\\
    -i k(\tau')&0\end{array}\,\right]     \label{matrixF}
    \end{eqnarray}
(note the position of time entries associated with the variables
$\varPhi=(\varphi(\tau),\pi)$ and $\varPhi'=(\varphi(\tau'),\pi)$).

The dependence of this determinant on $g(\tau)$ and $k(\tau)$ can be found from its functional variation with respect to these functions. From (\ref{I2}) we have
    \begin{eqnarray}
    &&\!\!\!\!\!\delta\ln\Big({\rm Det_*}
    \mbox{\boldmath${F}$}\Big)
    =-2\delta\ln Q+2\delta\ln ||g||
    +{\rm Tr}\,\Big(\delta\mathbb{F}\,
    \mathbb{G}\Big),                         \label{var1}
    \end{eqnarray}
where $\mathbb{G}$ is the Green's function of $\mathbb{F}$,
$\mathbb{F}\,\mathbb{G}=\mathbb{I}$ and the functional trace of any matrix with the block-structure of (\ref{matrixF}) is defined as
    \begin{eqnarray}
    &&{\rm Tr}
    \left[\,\,\begin{array}{cc}A(\tau,\tau')\,&\,\,\,
    B(\tau)\,\\
    B(\tau')&a\end{array}
    \,\right]=\oint d\tau\,A(\tau,\tau)+a\,.  \nonumber
    \end{eqnarray}

The block structure of the
matrix Green's function $\mathbb{G}$ has the form
    \begin{eqnarray}
    \mathbb{G}=
    \left[\,\,\begin{array}{cc}G(\tau,\tau')\,&\,\,\,
    {\displaystyle \frac{\textstyle i g(\tau)}Q}\,\\
    {\displaystyle \frac{\textstyle i g(\tau')}Q}
    &0\end{array}
    \,\right],                                    \label{matrixG}
    \end{eqnarray}
where the Green's function $G(\tau,\tau')$ in the diagonal block
satisfies the system of equations
    \begin{eqnarray}
    &&{\mbox{\boldmath$F$}}\,
    G(\tau,\tau')=\delta(\tau,\tau')
    -\frac{k(\tau)\,g(\tau')}Q,            \label{Gequation}\\
    &&\oint d\tau\,k(\tau)\,G(\tau,\tau')
    =0,                                           \label{Ggauge}
    \end{eqnarray}
which uniquely fix it. The second equation imposes the needed gauge,
whereas the right hand side of the first equation implies that
$G(\tau,\tau')$ is the inverse of the operator $F$ on the subspace
orthogonal to its zero mode.

The trace of the functional block-structure matrix in (\ref{var1}) corresponding to the variation of $g(\tau)$ reads
    \begin{equation}
    {\rm Tr}\,\Big(\delta_g\mathbb{F}\,
     \mathbb{G}\Big)=
     \mathrm{Tr}\Big(\delta\mbox{\boldmath$F$}\,
     G(\tau,\tau')\Big)\equiv\oint d\tau\,
    \delta\mbox{\boldmath$F$}\,
    G(\tau,\tau')\Big|_{\,\tau'=\tau}.      \label{deltagFG}
    \end{equation}
A similar variation of the gauge-fixing function gives a vanishing answer
    \begin{eqnarray}
    &&\!\!\!\!\!\!\delta_k\ln\Big({\rm Det_*}\,
    \mbox{\boldmath$F$}\Big)=
    -\frac2Q\oint d\tau\,\delta k\, g\nonumber\\
    &&\!\!\!\!\!\!
    +{\rm Tr}\!\left[\begin{array}{cc}{\displaystyle\frac1Q}\,{ g(\tau)\,\delta k(\tau')}\,&\,\,\,
    0\,\\
    { i\oint dy\,\delta k(y)\,
    G(y,\tau')}\,\,\,
    &{\displaystyle\frac1Q}
    {\oint dy\,
    \delta k(y)\,g(y)}\end{array}
    \right]=0,                     \label{gaugeindependence}
    \end{eqnarray}
as, of course, it should be in view of the gauge independent nature
of the Faddeev-Popov path integral. This guarantees the uniqueness of the definition of the reduced determinant ${\rm Det_*}\,\mbox{\boldmath$F$}$.

\section{Periodic Green's function}
For the calculation of the variation (\ref{deltagFG}) above we need the Green's function of the problem (\ref{Gequation})-(\ref{Ggauge}) which should be periodic on the circle (\ref{circle}). This means that the value of $G(\tau,\tau')$ and its derivative with respect to $\tau$ should match at $\tau=\tau_0+0$ and $\tau=\tau_0+T-0$. To achieve this property we will slightly extend the circle domain to the left of the point $\tau_0$
    \begin{eqnarray}
    \tau_0-\varepsilon<\tau<\tau_0+T,
    \quad\varepsilon>0,          \label{circle1}
    \end{eqnarray}
with an arbitrarily small positive $\varepsilon$ and demand that the monodromy of $G(\tau,\tau')$ should be vanishing for this small $\varepsilon$-range of $\tau$ near $\tau_0$
    \begin{equation}
    {\hat{\Delta}_{T}}G(\tau,\tau')\equiv G(\tau+T,\tau')-G(\tau,\tau')=0,\;
    \tau_0-\varepsilon<\tau<\tau_0.       \label{MonodromyG}
    \end{equation}

The ansatz for $G(\tau,\tau')$ can be as usual built with the aid of two linearly independent basis functions of the operator. One basis function coincides with the periodic zero mode $g(\tau)$ and another one $\psi(\tau)$ can be composed of the set of functions $\varPsi_i(\tau)$ defined by (\ref{Psis}) on various segments $\tau_{i-1}<\tau<\tau_i$.\footnote{For the extended range (\ref{circle1}) the missing $\varPsi_0(\tau)$ can be defined by identifying $\tau_{-1}$ with $\tau_{2k-1}-T$ and choosing some $\tau_0^*$ in $\tau_{-1}<\tau_0^*<\tau_0$.} For an arbitrary choice of auxiliary points  $\tau_i^*$ in (\ref{Psis}) the composite function
    \begin{eqnarray}
    &&\psi(\tau)=\varPsi_i(\tau),\;
    \qquad \tau_{i-1}\leq\tau\leq\tau_i,     \label{psi}
    \end{eqnarray}
will be continuous in view of (\ref{psilimits}), but the continuity of its derivative will generally be broken, because generally the equality $\dot\varPsi_i(\tau_i-0)=\dot\varPsi_{i+1}(\tau_i+0)$ is not satisfied. However, this equality for $i=1,2,...,2k-1$ can be enforced by a special choice of these auxiliary points $\tau_i^*$, becoming the equation for their determination. The solution for $\tau_i^*$ is unique, always exists and belongs to the corresponding segment $\tau_{i-1}<\tau_i^*<\tau_i$.\footnote{Indeed, the quantity $d\dot\varPsi_i(\tau_i)/d\tau_i^*=-\dot g(\tau_i)/g^2(\tau_i^*)$ is a sign definite function of $\tau_i^*$ nowhere vanishing on this segment, its absolute value quadratically divergent to $\infty$ at its boundaries. This in its turn means that $\dot\varPsi_i(\tau_i)$ is a monotonic function of $\tau_i^*$ which also ranges between $-\infty$ and $+\infty$ and therefore guarantees the unique solution for $\tau_i^*$ on this segment.} On the other hand, the continuity of the derivative of $\psi(\tau)$ cannot be attained at all roots of $g(\tau)$, $i=1,2,...2k$, because it would correspond to the existence of the second zero mode periodic on the circle, which is ruled out by construction. Therefore, the second basis function of $\mbox{\boldmath$F$}$ is not periodic on the circle, but in view of periodicity of the operator it satisfies the fundamental monodromy property (\ref{Monpsi}). Below we construct the periodic Green's function of $\mbox{\boldmath$F$}$ in terms of the monodromy parameter $\Delta$.

The Green's function of the problem  (\ref{Gequation})-(\ref{Ggauge}) can be represented as a sum of the particular solution of the inhomogeneous equation (\ref{Gequation}) and the bilinear combination of the basis functions $g$ and $\psi$ with the coefficients providing the periodicity property (\ref{MonodromyG}). In view of the Hermiticity of the operator $\mbox{\boldmath$F$}$ the kernel of its Green's is symmetric, $G(\tau,\tau')=G(\tau',\tau)$, so that we will look for it in the form
    \begin{eqnarray}
     &&G(\tau,\tau') = G_F(\tau,\tau') + \frac1Q\, \Omega(\tau,\tau')
     +\alpha\, H_{\psi\psi}(\tau,\tau')\nonumber\\
     &&\qquad\qquad+ \beta\,H_{\psi g}(\tau,\tau')
     +\gamma\,H_{gg}(\tau,\tau')\;,    \label{G1}
    \end{eqnarray}
where
    \begin{eqnarray}
     &&G_F(\tau,\tau')
     \equiv\frac12\big(g(\tau)\psi(\tau')
     -\psi(\tau)g(\tau')\big)\;\theta(\tau -\tau')\nonumber\\
     &&\qquad\quad
     +\frac12\big(\psi(\tau)g(\tau')
     -g(\tau)\psi(\tau')\big)\;
     \theta(\tau'-\tau\! )\;,                    \label{G_F}\\
     &&\Omega(\tau,\tau') \,\equiv\,
     \omega(\tau)\,g(\tau')
     +g(\tau)\,\omega(\tau')\;,                 \label{G_Omega}\\
     &&H_{\psi\psi}(\tau,\tau')\,
     \equiv\,\psi(\tau)\psi(\tau')\;,            \label{H1}\\
     &&H_{\psi g}(\tau,\tau') \,\equiv\,
     \psi(\tau)\,g(\tau')+
     g(\tau)\,\psi(\tau')\;,                    \label{H2}\\
     &&H_{gg}(\tau,\tau') \,
     \equiv\, g(\tau)\,g(\tau')\;,              \label{H3}
    \end{eqnarray}
and the function $\omega(\tau)$ will be defined below. The first term of (\ref{G1}) generates the delta-function in the right hand side of the equation (\ref{Gequation}), the second term $\Omega(\tau,\tau')/Q$ gives $-k(\tau)\,g(\tau')/Q$, while $H_{\psi\psi}(\tau,\tau')$, $H_{\psi g}(\tau,\tau')$ and $H_{gg}(\tau,\tau')$ represent symmetric solutions of the homogeneous equation with coefficients $\alpha$, $\beta$ and $\gamma$ to be fixed by the periodicity condition and the gauge condition (\ref{Ggauge}). The concrete form of these functions can be explained as follows.

The simplest form of a particular solution of $\mbox{\boldmath$F$}G_F(\tau,\tau')=\delta(\tau,\tau')$ can be written down with the aid of the Heviside step function as the combination $g(\tau)\psi(\tau')\;\theta(\tau -\tau')+(\tau\leftrightarrow\tau')$ bilinear in two basis functions $g$ and $\psi$ (normalized by the condition $W[g,\psi]=1$). But we will be interested in the kernel vanishing at the coincident arguments, which is easily attained by adding a special solution of the homogeneous equation,
    \begin{eqnarray}
     &&G_F(\tau,\tau')
     =g(\tau)\psi(\tau')\;\theta(\tau -\tau')
     +\psi(\tau)g(\tau')\;\theta(\tau'-\tau\!)\nonumber\\
     &&\qquad\quad-
     \frac12\,\big(g(\tau)\psi(\tau')
     +\psi(\tau)g(\tau')\big).   \nonumber
     \end{eqnarray}
This leads to (\ref{G_F}).

A particular solution of $\mbox{\boldmath$F$}\Omega(\tau,\tau')=-k(\tau)\,g(\tau')$ has the form (\ref{G_Omega}) where the function $\omega(\tau)$ satisfies the equation $\mbox{\boldmath$F$}\omega(\tau)=-k(\tau)$. It has as the first integral
    \begin{eqnarray}
     &&W[g,\omega](\tau) = Q(\tau)\;,     \nonumber\\
     &&Q(\tau)\equiv
    \int_{\tau_*}^{\tau} dy\, g(y)\,k(y), \label{Qtau}
    \end{eqnarray}
with an arbitrary $\tau_*$. This can be further integrated with a particular boundary condition $\omega(\tau_*) =0$ to give the final expression for $\omega(\tau)$,
    \begin{eqnarray}
     &&\omega(\tau)=  \psi (\tau) \int\limits_{\tau_\star}^{\tau} dy \,g(y)k(y) - g(\tau) \int\limits_{\tau_{\ast}}^{\tau} dy \,\psi(y)k(y)\;\nonumber\\
     &&\qquad\qquad=\psi (\tau) \, Q(\tau) - g(\tau) \int\limits_{\tau_*}^{\tau}
     dy \,\psi(y)\,k(y)\;.           \label{def_omega_ast_st}
    \end{eqnarray}

Similarly to $\psi(\tau)$ this function has a nontrivial monodromy which can be derived from the monodromy of $\psi(\tau)$,
    \begin{eqnarray}\label{omega_Mon_Derivation}
     &&{\hat{\Delta}_{_T}} \omega(\tau)=\psi (\tau{+}T)\!\! \int\limits_{\tau_*}^{\tau+T}\!\! dy\,g(y)\,k(y)
        -\psi (\tau)\!\int\limits_{\tau_*}^{\tau}dy \,g(y)\,k(y)\nonumber\\
        &&\qquad\qquad\qquad-g(\tau) \left(\int\limits_{\tau_*}^{\tau+T}-\int\limits_{\tau_*}^{\tau}\,\right) dy\,\psi(y)\,k(y) \nonumber\\
     &&\qquad\qquad=\psi (\tau{+}T)\,Q
        + \Delta\, g(\tau)\! \int\limits_{\tau_*}^{\tau}dy\, g\,k\nonumber\\
        &&\qquad\qquad\qquad
        -g(\tau) \!\!\int\limits_{\tau_*}^{\tau_*+T}\!\! dy\,\psi\,k- g(\tau) \!\int\limits_{\tau_*}^{\tau} dy\,\Big(\psi+ \Delta\, g\Big)\,k\nonumber\\
        &&\qquad\qquad\qquad+ g(\tau) \int\limits_{\tau_*}^{\tau}dy\, \psi\,k \nonumber\\
     &&\qquad\qquad=\psi(\tau+T) \,Q   -   g(\tau) \,\int\limits_{\tau_{*}}^{\tau_{*}+T} \!\!dy\,\psi(y)\,k(y)\;.
    \end{eqnarray}
Thus finally
    \begin{equation}
     {\hat{\Delta}_{_T}} \omega(\tau)\,
     = \psi(\tau)\,Q+g(\tau)
     \left(-\!\!\int_{\tau_*}^{\tau_*+T}\!\! dy \,\psi(y)\,k(y)+\Delta\,Q\right).    \label{Monomega}
    \end{equation}

Using this monodromy and (\ref{Monpsi}) we have
    \begin{eqnarray}
    &&{\hat{\Delta}_{_T}}\, G_F(\tau,\tau')
    =\frac12\Big(\,g(\tau+T)\psi(\tau')
    -\psi(\tau+T)g(\tau')\,\Big)\nonumber\\
    &&\qquad\qquad\qquad -\frac12\Big(\,\psi(\tau)g(\tau')-g(\tau)\psi(\tau')\,\Big)
     \nonumber\\
     &&\quad\quad
     =g(\tau)\,\psi(\tau')-\psi(\tau)\,g(\tau')-\frac12\,\Delta \,g(\tau)\,g(\tau')\;,
    \end{eqnarray}
where we took into account that for the range of $\tau$ in the small $\varepsilon$-domain of the enlarged range (\ref{circle1}), $\tau_0-\varepsilon<\tau\leq\tau_0$ and $\tau_0\leq\tau'<\tau_0+T$ the arguments of $G_F(\tau,\tau')$ satisfy the relations $\tau<\tau'$ and $\tau+T>\tau'$ (the latter inequality can always be enforced by choosing a small positive $\varepsilon$ in the range $0<\varepsilon<\tau_0+T-\tau'$). Similarly
    \begin{eqnarray}
     &&{\hat{\Delta}_{_T}}\,\Omega(\tau,\tau')\,
     ={\hat{\Delta}_{_T}} \omega(\tau)\, g(\tau')
     =Q\,\psi(\tau)\,g(\tau')\nonumber\\
     &&\qquad\quad
     -g(\tau)\,g(\tau')
     \int_{\tau_{*}}^{\tau_{*}+T} \!\!dy\, \psi(y)\, k(y)
        + Q\,\Delta\; g(\tau)\,g(\tau')\;,
     \nonumber\\
     &&{\hat{\Delta}_{_T}} H_{\psi\psi}(\tau,\tau')
     =\Delta\,g(\tau)\,\psi(\tau')\;,
     \nonumber\\
     &&{\hat{\Delta}_{_T}} H_{\psi g}(\tau,\tau')
     =\Delta\,g(\tau)\,g(\tau')\;;
     \nonumber\\
     &&{\hat{\Delta}_{_T}} H_{gg}(\tau,\tau')\,
     =\,0\;,                       \label{Gk_Monodromy_Prop}
    \end{eqnarray}
so that the periodicity equation (\ref{MonodromyG}) takes the form
    \begin{eqnarray}
     &&{\hat{\Delta}_{T}} G(\tau,\tau')
     =g(\tau)\,\psi(\tau')\,\Big(\alpha\,
     \Delta + 1\Big)
     \nonumber\\
     &&\qquad\quad
     +g(\tau)\,g(\tau')\,\Big(\,\frac{\Delta}2
     +\beta\,\Delta -\frac1Q\!\int_{\tau_{*}}^{\tau_{*}+T}
     \!\!dy\, \psi(y)\,k(y)\Big)\nonumber\\
     &&\qquad\quad=0,
    \end{eqnarray}
whence
    \begin{equation}
     \alpha = - \frac1{\Delta}\,,\qquad
     \beta = -\frac12 \,+\, \frac1{\Delta \,Q}\!\int\limits_{\tau_{*}}^{\tau_{*}+T}
     \!\!dy\, \psi(y)\, k(y).                    \label{ab}
    \end{equation}

Finally, the periodicity requirement does not impose any restriction on the coefficient $\gamma$, but it immediately follows from the gauge fixing condition (\ref{Ggauge}),
    \begin{eqnarray}
    &&\gamma = \frac1Q \int\limits_{\tau_{*}}^{\tau_{*}+T} \!\!dy\, \psi(y)\, k(y)-\frac1{Q^2 \Delta} \left(\int\limits_{\tau_{*}}^{\tau_{*}+T} \!\!dy\, \psi(y)\, k(y)\right)^2\nonumber\\
     &&\qquad\qquad\qquad\qquad
     - \frac1{Q^2}\int\limits_{\tau_{*}}^{\tau_{*}+T} \!\!dy\, \omega(y)\, k(y).
    \end{eqnarray}

\section{Variation of the determinant}
The calculation of the variational term (\ref{deltagFG}) is based on integration by parts and a systematic use of the Wronskian relation (\ref{Wronskianrelation}) together with equations for $g$ and $\psi$. Using the obtained Green's function (\ref{G1}) we see that various terms of (\ref{deltagFG}) transform as follows.

The first term contributes zero because the coincidence limit $G_F(\tau,\tau)$ is vanishing,
    \begin{eqnarray}
     &&\mathrm{Tr} \Big(\delta\mbox{\boldmath$F$}\,G_F(\tau,\tau')\Big)
     \equiv \oint d\tau\;
     \delta \overrightarrow{\mbox{\boldmath$F$}}\,
     G_F(\tau,\tau')\big|_{\,\tau'=\tau}\nonumber\\
     &&\qquad\qquad\qquad=\oint d\tau\;\,\delta \!\!\left(\frac{\ddot{g}}{g}\right)\,G_F(\tau,\tau)= 0.
    \end{eqnarray}
The contribution of the rest of the terms of (\ref{G1}) represent integrals over the period $T$ of non-periodic functions, which explicitly depend on the choice of the initial integration point in
    \begin{eqnarray}
    \oint d\tau\,(...)=\int_{\tau_-}^{\tau_-+T}d\tau\,(...).
    \end{eqnarray}
To check that in the final answer this dependence cancels out, as it should for the full periodic function $\delta{\mbox{\boldmath$F$}}\,G_F(\tau,\tau)$, we keep $\tau_-$ arbitrarily fixed in what follows. Thus, the contribution of the $\Omega$-term transforms as
    \begin{eqnarray}
     &&\mathrm{Tr}\Big(\delta\mbox{\boldmath$F$}\;
     \Omega(\tau,\tau')\Big)
       = \int_{\tau_-}^{\tau_-+T}d\tau\; \Big(\omega\,\delta\overrightarrow{\mbox{\boldmath$F$}} \,g+g\,\delta\overrightarrow{\mbox{\boldmath$F$}} \,\omega\Big)\nonumber\\
      &&\qquad\qquad\qquad=- \int_{\tau_-}^{\tau_-+T}d\tau\; \Big(\omega\, \overrightarrow{\mbox{\boldmath$F$}} \,\delta g\,
      +\,g\,\overrightarrow{\mbox{\boldmath$F$}}\,\delta \omega\Big)
     \nonumber\\
     &&\qquad\qquad\qquad
      =-\! \int_{\tau_-}^{\tau_-+T}d\tau\; \Big(\omega\, \overleftarrow{\mbox{\boldmath$F$}}\,\delta g\,
      +\,g\,\overleftarrow{\mbox{\boldmath$F$}} \,\delta_g \omega\Big)\nonumber\\
      &&\qquad\qquad\qquad\qquad
      +\,\Big(W[\omega,\delta g]
      +W[g,\delta\omega]\Big)\Big|_{\,\tau_-}^{\,\tau_-+T}
     \nonumber\\
     &&\qquad\qquad
      =\int_{\tau_-}^{\tau_-+T}\!\!\!d\tau\:  k \,\delta g\,
      +{\hat{\Delta}_{_T}}W[\omega,\delta g](\tau_-)\nonumber\\
      &&\qquad\qquad\qquad\qquad\qquad+\,{\hat{\Delta}_{_T}}W[g,\delta_g \omega](\tau_-).
    \end{eqnarray}
where we used the equations $\mbox{\boldmath$F$}g=0$ and $\overrightarrow{\mbox{\boldmath$F$}}\omega=-k$ and their variational corollaries $\big(\delta\overrightarrow{\mbox{\boldmath$F$}}\big) \,g=-\overrightarrow{\mbox{\boldmath$F$}}\,\delta g$ and $\big(\delta\overrightarrow{\mbox{\boldmath$F$}}\big)\omega=
-\overrightarrow{\mbox{\boldmath$F$}}\delta_g\omega-\delta_g k=
-\overrightarrow{\mbox{\boldmath$F$}}\delta_g\omega$. Similarly, the contributions of $H_{\psi\psi}$, $H_{\psi g}$ and $H_{gg}$ terms read
    \begin{eqnarray}
    &&\mathrm{Tr}\Big((\delta\mbox{\boldmath$F$})\; H_{\psi\psi}(\tau,\tau')\Big)
    =\int_{\tau_-}^{\tau_-+T}d\tau\; \psi\,
    \delta\overrightarrow{\mbox{\boldmath$F$}} \,\psi\nonumber\\
    &&\qquad=-\int_{\tau_-}^{\tau_-+T}d\tau\; \psi\,
    \overrightarrow{\mbox{\boldmath$F$}}\,\delta\psi
    ={\hat{\Delta}_{_T}}
    W[\psi,\delta\psi](\tau_-)\;,\\
    &&\mathrm{Tr}\Big(\delta\mbox{\boldmath$F$}\;
    H_{\psi g}(\tau,\tau')\Big)
     = \int_{\tau_-}^{\tau_-+T}d\tau\; \Big(\psi\,\delta\overrightarrow{\mbox{\boldmath$F$}} \,g+g\,\delta\overrightarrow{\mbox{\boldmath$F$}} \,\psi\Big)\nonumber\\
      &&\qquad\qquad=- \int_{\tau_-}^{\tau_-+T}d\tau\; \Big(\psi\, \overrightarrow{\mbox{\boldmath$F$}} \,\delta g\,
      +\,g\,\overrightarrow{\mbox{\boldmath$F$}}\,\delta \psi\Big)
      \nonumber\\
      &&\qquad\qquad
      ={\hat{\Delta}_{_T}}\big( W[\psi,\delta g](\tau_-)
      + W[g,\delta\psi](\tau_-)\Big)\;,\\
    &&\mathrm{Tr}\Big(\delta\mbox{\boldmath$F$}\; H_{gg}(\tau,\tau')\Big)
    =\int_{\tau_-}^{\tau_-+T}d\tau\; g\,
    \delta\overrightarrow{\mbox{\boldmath$F$}}\,g\nonumber\\
    &&\quad=-\int_{\tau_-}^{\tau_-+T}d\tau\; g\,
    \overrightarrow{\mbox{\boldmath$F$}}\,\delta g
    ={\hat{\Delta}_{_T}}
    W[g,\delta g](\tau_-)=0\;.
    \end{eqnarray}
Therefore, the variation (\ref{deltagFG}) takes the following form in terms of the monodromies of the set of Wronskians
    \begin{eqnarray}
     &&\mathrm{Tr}\Big(\delta\mbox{\boldmath$F$}\,
     G(\tau,\tau')\Big)=
     \frac1Q \int\limits_{\tau_-}^{\tau_-+T}
     \!\!\!d\tau\,k\,\delta g\nonumber\\
     &&\qquad+\frac1Q\,\Big( \hat{\Delta}_{_T}W[\omega,\delta g](\tau_-)
      + \hat{\Delta}_{_T}W[g,\delta\omega](\tau_-)\Big)
     \nonumber\\
     &&\qquad
     +\alpha\,{\hat{\Delta}_{_T}} W[\psi,\delta\psi](\tau_-)\nonumber\\
     &&\qquad+\beta\,\Big( {\hat{\Delta}_{_T}}W[\psi,\delta g](\tau_-)
     +{\hat{\Delta}_{_T}}
     W[g,\delta\psi](\tau_-)\Big).               \label{trace1}
    \end{eqnarray}

These monodromies in their turn express via the monodromy parameter $\Delta$ of the function $\psi(\tau)$ and its variation. First, in view of the relations $W[g,g\,]=0$ and $W[\psi,g]=-1$ we have
    \begin{eqnarray}
     &&{\hat{\Delta}_{_T}}\, W[\psi,\delta \psi](\tau_-)
     = W[\big(\psi+\Delta\,g\big),\delta \big(\psi+\Delta\, g\big)](\tau_-)\nonumber\\
     &&\qquad\qquad\qquad\qquad-W[\psi,\delta\psi](\tau_-)
      \nonumber\\
       &&\qquad\qquad\qquad=
      -\delta\Delta
      +\Delta\, W[\psi,\delta  g]
      +\Delta\, W[g,\delta \psi]\nonumber\\
       &&\qquad\qquad\qquad
      +\Delta^2 \,W[ g,\delta g].         \label{Wpsipsi2}
    \end{eqnarray}
Second, using the relation $W[g,\delta\psi]=\delta W[g, \psi]
-W[\delta g, \psi]$ we have the chain of identities for the
    \begin{eqnarray}
     &&{\hat{\Delta}_{_T}}\Big( W[\psi,\delta g]
     +W[g,\delta\psi]\Big)(\tau_-)
     \nonumber\\
     &&\qquad\qquad
     ={\hat{\Delta}_{_T}}\big( W[\psi,\delta g]
      + \delta W[g, \psi]
      - W[\delta g, \psi]\big)(\tau_-)
     \nonumber\\
     &&\qquad\qquad=2 W[\big(\psi
     +\Delta g\big),\delta g](\tau_-)
     -2 W[\psi,\delta g](\tau_-)
    \nonumber\\
     &&\qquad\qquad
     = 2\Delta W[g,\delta g](\tau_-),              \label{Wgpsi}
    \end{eqnarray}
where we used the fact that $W[g,\psi]=1$ and $\delta W[g,\psi]=0$. Similarly
    \begin{eqnarray}
     &&\hat{\Delta}_{_T}\Big( W[\omega,\delta g]+W[g,\delta_g \omega]\Big)(\tau_-)\nonumber\\
     &&\qquad\qquad
     ={\hat{\Delta}_{_T}} \Big( W[\omega,\delta g]
     +\delta W[g,\omega]-W[\delta g,\omega]\Big)(\tau_-)
     \nonumber\\
     &&\qquad\qquad
     =2 W[{\hat{\Delta}_{_T}}\omega,\delta g](\tau_-) +\delta\hat{\Delta}_{_T}Q(\tau_-),     \nonumber
     \end{eqnarray}
because $W[\omega,g](\tau)=Q(\tau)$ and ${\hat{\Delta}_{_T}}\delta Q=\delta\hat{\Delta}_{_T}Q$. Finally, we use the monodromy of $\omega$ given by Eq.(\ref{Monomega}) and the fact that the monodromy of the function $Q(\tau)$ defined by (\ref{Qtau}) equals $\hat{\Delta}_{_T}Q(\tau_-)=Q$ where $Q$ is defined by Eq.(\ref{Q}). Thus we have for this pair of monodromies
    \begin{eqnarray}
     &&\hat{\Delta}_{_T}\Big( W[\omega,\delta g] +W[g,\delta\omega]\Big)(\tau_-)
     =2\,Q\,W[\psi(\tau),\,\delta g](\tau_-)
     \nonumber\\
     &&
     +2\Big(\Delta\,Q-\int_{\tau_{*}}^{\tau_{*}+T}
     \!\!dy\,\psi(y)\, k(y)\Big)\,
     W[g,\delta g] (\tau_-)
     +\delta Q.                        \label{Wgomega}
    \end{eqnarray}

Using in (\ref{trace1}) the expressions (\ref{ab}) for $\alpha$ and $\beta$ and the above set of Wronskian monodromies (\ref{Wpsipsi2})-(\ref{Wgomega}) we eventually find
    \begin{eqnarray}
     &&\mathrm{Tr} \Big(\delta\mbox{\boldmath$F$}\,G(\tau,\tau')\Big)
     =2\,\frac{\delta Q}Q + \frac{\delta\Delta}{\Delta}
     + W[\psi,\delta  g](\tau_-)\nonumber\\
     &&\qquad\qquad\qquad\qquad
     - W[g, \delta \psi](\tau_-)\nonumber\\
     &&\qquad\qquad\qquad\qquad
     =2\,\delta\ln Q + \delta\ln\Delta\,.    \label{varTr_summary}
    \end{eqnarray}
Here the $\tau_-$-dependent terms  proportional to $W[g,\delta g](\tau_-)$ and in $W[\psi,\delta  g](\tau_-) - W[g, \delta \psi](\tau_-)=\;\;\delta W[\psi,g]=0$ dutifully cancel out. Therefore, the total variation (\ref{var1}) becomes independent of $\tau_-$ and gauge independent (as it should, c.f. Eq.(\ref{gaugeindependence}))
    \begin{eqnarray}
    \delta\ln\Big({\rm Det_*}\,
    \mbox{\boldmath${F}$}\Big)
    =2\,\delta\ln ||g||+\delta\ln\Delta,
    \end{eqnarray}
and we finally get the explicit answer for ${\rm Det_*}\,\mbox{\boldmath${F}$}$
    \begin{eqnarray}
    {\rm Det_*}\,
    \mbox{\boldmath${F}$}
    =C(T)\times
    \Delta\oint d\tau\,g^2(\tau)\,.    \label{det1}
    \end{eqnarray}

This result is valid up to an overall coefficient $C(T)$ which is functionally independent of $g(\tau)$, but can depend on the period $T$ -- the only remaining free parameter. This coefficient function can be determined by the zeta-function method for a particular case of the constant function $g(\tau)=c$ corresponding to the operator $\mbox{\boldmath${F}$}=-d^2/d\tau^2$ with the explicit spectrum of eigenfunctions and respective eigenvalues
    \begin{eqnarray}
    &&\varphi_0=1,\quad \lambda_0=0,\nonumber\\
    &&\varphi_{1\,n}(\tau)
    =\sin\big(2\pi n\tau/T\big),\; \varphi_{2\,n}(\tau)
    =\cos\big(2\pi n\tau/T\big),\nonumber\\
    &&\lambda_n=\big(2\pi n/T\big)^2,
    \quad n=1,2,...\;.
    \end{eqnarray}
The logarithm of the corresponding restricted determinant -- the product of all nonvanishing eigenvalues regularized by zeta-function method -- equals
    \begin{eqnarray}
    &&\ln{\rm Det_*}\!\left(-\frac{d^2}{d\tau^2}\right)
    =2\sum\limits_{n=1}^\infty
    \ln\left(\frac{2\pi n}{T}\right)^2\nonumber\\
    &&\qquad\quad=4\ln\left(\frac{2\pi}{T}\right)
    \zeta_R(0)-4\zeta'_R(0)=2\ln T.   \label{1000}
    \end{eqnarray}
Here $\zeta_R(s)=\sum_{n=1}^\infty n^{-s}$ is the Riemann zeta function having the following particular value $\zeta_R(0)=-\frac12$ and the value of its derivative $\zeta_R'(0)=-\frac12\ln{2\pi}$. On the other hand, the basis functions and the monodromy $\Delta$ for this operator read
    \begin{eqnarray}
    &&g(\tau)=c,\quad \oint d\tau\,g^2=c^2\,T,\nonumber\\
    &&\psi(\tau)=\frac1c(\tau-\tau_*),\quad\Delta=\frac{T}{c^2}.
    \end{eqnarray}
Therefore according to (\ref{det1}) ${\rm Det_*}\!\left(-d^2/d\tau^2\right)=C(T)\,T^2$, and the comparison with (\ref{1000}) gives the $T$-independent result $C(T)=1$, which confirms the unit normalization coefficient in (\ref{det}).

\section{The monodromy algorithm}
From the definition of the monodromy parameter (\ref{Monpsi}) it follows that
    \begin{eqnarray}
    \Delta=\frac{\psi(\tau+T)-\psi(\tau)}{g(\tau)}.
    \end{eqnarray}
By tending the point $\tau$ to the one of the roots of $g(\tau)$, $\tau\to\tau_0$, $\tau+T\to\tau_{2k}$, we have
    \begin{eqnarray}
    &&\Delta=\frac{\dot\psi(\tau_{2k})-\dot\psi(\tau_0)}{\dot g(\tau_0)}=\left(\frac{\dot\psi(\tau_{2k})}{\dot g(\tau_{2k})}-
    \frac{\dot\psi(\tau_{2k-1})}{\dot g(\tau_{2k-1})}\right)+...\nonumber\\
    &&\qquad\qquad\qquad\qquad\quad+\left(
    \frac{\dot\psi(\tau_1)}{\dot g(\tau_1)}-
    \frac{\dot\psi(\tau_0)}{\dot g(\tau_0)}\right),
    \end{eqnarray}
where we took into account that $\dot g(\tau_{2k})=\dot g(\tau_0)$. Thus we immediately arrive at the algorithm (\ref{Delta})-(\ref{Deltai}). It is important that unlike in the construction of the function $\psi(\tau)$ which has to be smooth on $S^1$ at all roots $\tau_i$ except $\tau_0$ (the property that was attained above by a special choice of the auxiliary points $\tau_i^*$), the derivatives of neighboring functions $\varPsi_i(\tau)$ in (\ref{Delta})-(\ref{Deltai}) should not necessarily be matched at these junction points. This is because the partial contributions $\Delta_i$ to the overall monodromy $\Delta$ are individually independent of  $\tau_i^*$,
    \begin{eqnarray}
    \frac{d\varDelta_i}{d\tau_i^*}=0,
    \end{eqnarray}
which can be easily verified by using a simple relation
$d\dot\varPsi_i(\tau)/d\tau_i^*=-\dot g(\tau)/g^2(\tau_i^*)$. Thus, the monodromy is uniquely defined and independent of the choice of the auxiliary points $\tau_i^*$ necessarily entering the definition of functions $\varPsi_i(\tau)$ in Eq.(\ref{Psis}).

To derive the manageable integral expression (\ref{derivative}) for $\dot\varPsi(\tau)$ at the roots $\tau_\pm$ of a zero mode, we note that its derivative
    \begin{eqnarray}
     &&\dot\varPsi(\tau)= \dot g(\tau)
     \int\limits_{\tau^*}^{\tau}
     \frac{dy}{g^2(y)}+\frac1{g(\tau)}
    \end{eqnarray}
has two terms which separately diverge when $\tau\to\tau_\pm$. To circumvent this difficulty one can add and subtract in the integrand of the first term the total derivative term in the integration variable $y$, $(1/\dot g(\tau))(d/dy)(1/g(y))$. This yields the representation for the time derivative,
    \begin{eqnarray}
    \dot\varPsi(\tau)=
    \int\limits_{\tau^*}^{\tau_\pm} dy\,
    \frac{\dot g(\tau)
    -\dot g(y)}{g^2(y)}+\frac1{g(\tau^*)},    \label{derivative1}
    \end{eqnarray}
which now allows one to take the limit $\tau=\tau_\pm$ directly in the integrand, integration remaining convergent because the value of the numerator function $\dot g(\tau_\pm)-\dot g(y)$ and its $y$-derivative, $\propto \ddot g(y)$, are vanishing at $y=\tau_\pm$ (cf. Eq.(\ref{ddotg})). This gives (\ref{derivative}).

\subsection{CFT driven cosmology: single node case vs multiple nodes}
CFT driven cosmology considered in \cite{slih,why,PIQC} is characterized by the Einstein-Hilbert action with matter sources in the form of many field species conformally coupled to metric. For a large number of these species the path integral (\ref{2}) in this model with a good accuracy generates the effective action (\ref{action1}) with the local Lagrangian of minisuperspace variables and the free energy of matter fields
    \begin{eqnarray}
    &&{\cal L}(a,a')=m_P^2\left\{-aa'^2
    -a+ \frac\Lambda3\, a^3\right.
    \nonumber\\
    &&\qquad\qquad\qquad\quad
    \left.+B\left(\frac{a'^2}{a}
    -\frac{a'^4}{6 a}
    +\frac1{2a}\right)\,
    \right\},                  \label{Lagrangian}\\
    &&F(\eta)=\pm\sum_{\omega}
    \ln\big(1\mp e^{-\omega\eta}\big).  \label{F}
    \end{eqnarray}
The Lagrangian ${\cal L}(a,a')$ includes the minisuperspace Einstein-Hilbert term, the contributions of the overall conformal anomaly of these fields ($B>0$ is the coefficient of its topological Gauss-Bonnet term) and their vacuum Casimir energy. Here $m_P=(3\pi/4G)^{1/2}$ and $\Lambda$ are the renormalized Planck mass and cosmological constant, $a'=\dot a/N$ and the free energy $F(\eta)$ is the function of $\eta$ -- the inverse of the effective temperature, which is given by the full period of the conformal time, see (\ref{action1}). This is a typical boson or fermion sum over field oscillators with energies $\omega$ on a unit 3-sphere.

Semiclassically the integral (\ref{1}) is dominated by the saddle
points --- cosmological instanton solutions of the effective Friedmann equation of the Euclidean gravity theory
    \begin{eqnarray}
    &&-\frac{a'^2}{a^2}+\frac{1}{a^2}
    -B \left(\,\frac{a'^4}{2a^4}
    -\frac{a'^2}{a^4}\right) =
    \frac\Lambda3+\frac{C}{ a^4},          \label{efeq}\\
    &&C = \frac{B}2+\frac1{m_P^2}
    \frac{dF(\eta)}{d\eta}.                \label{bootstrap}
    \end{eqnarray}
This Friedmann equation is modified by the quantum $B$-term and contains the radiation term $C/a^4$, the constant $C$ characterizing the sum of the renormalized Casimir energy $B/2$ and the energy of the gas of thermally excited particles $dF(\eta)/d\eta=\sum_\omega\omega/(e^{\omega\eta}\mp1)$.

As shown in \cite{slih}, these solutions have a $k$-fold nature (with any integer $k=1,2,3,...$). This means that during the full period of its Euclidean time $T$ the scale factor oscillates $k$ times back and forth between its maximum and minimum values $a_\pm=a(\tau_\pm)$, $a_-\leq a(\tau)\leq a_+$,
    \begin{eqnarray}
    a^2_\pm=
    \frac3{2\Lambda}\big(\,1\pm\sqrt{1-4\Lambda C/3}\,\big), \label{apm}
    \end{eqnarray}
and forms a kind of a garland of $S^1\times S^3$ topology with
the oscillating $S^3$ section. Therefore, the inverse temperature $\eta$ -- the instanton period in units of the conformal time -- is given by the integral
    \begin{eqnarray}
    \eta=\int\limits_{\tau_0}^{\tau_0+k\mbox{\boldmath${T}$}} d\tau\,\frac{N}{a}
    =2k\int\limits_{\tau_-}^{\tau_+}d\tau\,\frac{N}{a}
    \end{eqnarray}
over the full period $T$ which is the $2k$-multiple of the integral between the two neighboring turning points $\tau_\pm$ of the scale factor history or $k$-multiple of one period $\mbox{\boldmath${T}$}$ of such an oscillation,
    \begin{eqnarray}
    \dot a(\tau_\pm)=0,\quad
    \mbox{\boldmath${T}$}=2(\tau_+-\tau_-),\quad T=k\mbox{\boldmath${T}$}.
    \end{eqnarray}
Thus, for $k>1$ the period $T$ of the periodic functions on the functional space of which we calculate the determinant ${\rm Det_*}\,\mbox{\boldmath${F}$}$ is the $k$-multiple of the actual period $\mbox{\boldmath${T}$}$ of the zero mode, so that the total period of the instanton is composed by glueing together $k$ such {\em fundamental} periods. Each of these periods contains two turning points of $a(\tau)$, and since these points coincide with the roots of the zero mode $g\propto\dot a$ this $k$-fold instanton gives rise to $k$ nodes of $g(\tau)$.

This situation is a particular case of our general setting of Sect.2, which corresponds to the ``equidistant" set of roots of the zero mode function (\ref{g}) associated with the turning points of the instanton solution. One can check that all the properties (\ref{roots})-(\ref{ddotg}) of the function $g(\tau)$ defined by Eq.(\ref{g}) for the Lagrangian (\ref{Lagrangian}) are satisfied\footnote{The last property (\ref{ddotg}) is guaranteed by the fact that for any solution of Eq.(\ref{efeq}) all odd order derivatives of $a(\tau)$ are vanishing at its turning points.}. Moreover, any solution $a(\tau)$ of this equation is symmetric under the reflection of the time variable relative to any of its turning points, $a(\tau_\pm+\tau)=a(\tau_\pm-\tau)$, so that $\dot a$ and $g\propto \dot a$ are antisymmetric under this reflection. Therefore, one root of $g$ can be placed right in the middle of the fundamental period of length $\mbox{\boldmath${T}$}$, so that all segments of time $[\tau_{i-1},\tau_i]$ connecting the pairs of neighboring roots are of the same length and $g(\tau)$ is antisymmetric under reflections with respect to each of these roots.

For the simplest single-node case with $k=1$ which was considered in \cite{PIQC,QCdet,oneloop1} the location of the roots can be parameterized as
    \begin{eqnarray}
    \tau_0=-\tau_+,\quad\tau_1=0,\quad\tau_2=\tau_+
    \equiv\frac12\,\mbox{\boldmath${T}$}
    \end{eqnarray}
with the points $\tau_0$ and $\tau_2$ identified (in order to make the time range, $-\tau_+<\tau<\tau_+$, to form the circle just as in (\ref{roots})). Then $g(\tau)$ is an odd function of $\tau$,
    \begin{eqnarray}
    &&g(\tau)=-g(-\tau),                    \label{conditionsong}
    \end{eqnarray}
and has two first degree zeros at antipodal points of this circle $\tau=\tau_-\equiv 0$ and $\tau=\pm\tau_+$ which mark the boundaries of two half-periods of the total time range, $-\mbox{\boldmath${T}$}/2<\tau<\mbox{\boldmath${T}$}/2$. Correspondingly the functions $\varPsi_i(\tau)$ defined by Eq.(\ref{Psis}) can be chosen in the form
    \begin{eqnarray}
    &&\varPsi_2(\tau)=\varPsi(\tau)\equiv
    g(\tau)\int\limits_{\tau_*}^{\tau}\frac{dy}{g^2(y)},\quad
    0<\tau<\tau_+,                         \label{Psi}\\
    &&\varPsi_1(\tau)=
    g(\tau)\int\limits_{-\tau_*}^{\tau}\frac{dy}{g^2(y)}
    =\varPsi(-\tau),\,
    -\tau_+<\tau<0,
    \end{eqnarray}
with some $\tau_*>0$ in the half period range of $\tau$. With this choice the monodromy constituents $\Delta_i$ become
    \begin{eqnarray}
    &&\Delta_1=\Delta_2=
    -\big(\,\varPsi_+\dot\varPsi_+
    -\varPsi_-\dot\varPsi_-\big),                 \label{bfI}\\
    &&\varPsi_\pm\equiv
    \varPsi(\tau_\pm),\,\,\,\,
    \dot\varPsi_\pm\equiv\dot\varPsi(\tau_\pm),    \label{Psipm}
    \end{eqnarray}
and the total monodromy reproduces the result obtained for a single-node case in \cite{QCdet} (denoted there by $\mbox{\boldmath${I}$}$)
    \begin{eqnarray}
    &&{\bf\Delta}\equiv\Delta\,\Big|_{\,k=1}=2\Delta_1=
    -2\,\big(\,\varPsi_+\dot\varPsi_+
    -\varPsi_-\dot\varPsi_-\big).              \label{bfDelta}
    \end{eqnarray}

Due to the ``equidistant" property of the set of roots discussed above all terms of the total monodromy (\ref{Delta}) coincide with (\ref{bfI}), $\Delta_i={\bf\Delta}/2$, and the monodromy for a multi-node background equals
    \begin{eqnarray}
    \Delta\,\Big|_{\,k>1}=k{\bf\Delta},      \label{k}
    \end{eqnarray}
where $\bf\Delta$ can be calculated by means of Eq.(\ref{bfDelta}) with the function $\varPsi(\tau)$ prescribed on the single half-period of the zero mode by Eq.(\ref{Psi}).\footnote{One should not think that the monodromy for the $k$-node case is numerically $k$ times larger than its value for the single-node one, because the histories $a(\tau)$ for multi-fold instantons are not obtained by simply matching together $k$ copies of the single-fold instanton solution. Simple additivity is violated in view of the ``bootstrap" equation (\ref{bootstrap}) relating the amount of radiation $C$ on the instanton background to its geometry. For this reason, in particular, the on-shell value of the instanton action is not a simple multiple of $k$, but rather tends to zero for $k\to\infty$ \cite{slih}.}

\section{Conclusions}
Thus we obtained the restricted functional determinant of a special second order operator (\ref{operator}), subject to periodic boundary conditions with a periodic zero mode $g=g(\tau)$. This operator $\mbox{\boldmath${F}$}$ determines the one-loop statistical sum for the microcanonical ensemble in cosmology generated by a conformal field theory. This ensemble realizes the concept of cosmological initial conditions by generalizing the notion of the no-boundary wavefunction of the Universe to the level of a special quasi-thermal state which is dominated by instantons with an oscillating scale factor of their Euclidean Friedmann-Robertson-Walker metric. These oscillations result in the multi-node nature of the zero mode $g(\tau)$, (\ref{g}), of $\mbox{\boldmath${F}$}$, which is gauged out in the path integral for the statistical sum by the method of the Faddeev-Popov gauge fixing procedure. This effectively leads to the restricted functional determinant (with the zero eigenvalue omitted).

In contrast to the previous work \cite{QCdet} we systematically used the monodromy method for the calculation of this determinant, which allowed us to generalize the previously known result for the single-node case \cite{QCdet} to the general case of multiple nodes (roots) of $g(\tau)$ within the instanton period of the Euclidean time $\tau$. The functional determinant of $\mbox{\boldmath${F}$}$ expresses in terms of the monodromy of its basis function (\ref{det}), which is obtained in quadratures as a sum of contributions of time segments connecting neighboring pairs of the zero mode roots within the period range (\ref{Delta})-(\ref{Deltai}). The advantage of this representation is that the solutions of homogeneous equation explicitly known in quadratures for these different segments (\ref{Psis}) should not necessarily match with one another to form a smooth basis function on the entire period, which leaves integration constants $\tau_i^*$ in (\ref{Psis}) arbitrary. This disentangles these segments and makes (\ref{Delta}) an additive sum of independent contributions.

For the single-node case of the CFT driven cosmology ($k=1$ number of oscillations of the background scale factor or $2k=2$ roots of the zero mode) we reproduce the result of \cite{QCdet}. Moreover, in view of the equidistant nature of the set of these roots for multi-node case, the monodromy becomes a $2k$-multiple of the contribution of the half-period of the zero mode $g(\tau)$ (\ref{bfI})-(\ref{k}). Application of this result for a single-fold instanton in \cite{oneloop1} showed smallness of its one-loop part in the statistical sum of the CFT driven cosmology. However, preliminary analysis for multiple-fold instantons shows the possibility of a phase transition, strong coupling domain and also peculiar behavior for $k\to\infty$. This limit corresponds to the upper bound of the admissible cosmological constant range $\Lambda_{\rm max}=3/2B$ -- a new quantum gravity scale which is of special interest in the microcanonical ensemble of the CFT driven cosmology \cite{slih,why}. Cosmological multi-fold instantons of \cite{slih,why} are analogous to multi-instanton solutions in Yang-Mills theory, but in contrast to the Yang-Mills case no dilute gas approximation applies here, and their action is not a multiple of the single-fold one and, moreover, tends to zero for $k\to\infty$ \cite{slih} (see footnote above). Thus, their contributions are not exponentially suppressed (or infinitely enhanced like in \cite{Halliwell-Myers} in view of the negative value of the classical gravitational action), and play important role in the full ensemble. Therefore, the above results for $k>1$ become indispensable for the calculation of their preexponential factors to be considered in \cite{multifold}.

\section*{Acknowledgements}
The authors are grateful to I.V.Tyutin and B.L.Voronov for helpful discussions. The work of D.V.N. was supported by the RFBR under Grant No. 11-02-00512 and the work of A.O.B. was partially supported by the RFBR under Grant No. 11-01-00830.

\end{document}